\documentclass{iaus}
\usepackage{graphics}

  \checkfont{eurm10}
  \iffontfound
    \IfFileExists{upmath.sty}
      {\typeout{^^JFound AMS Euler Roman fonts on the system,
                   using the 'upmath' package.^^J}%
       \usepackage{upmath}}
      {\typeout{^^JFound AMS Euler Roman fonts on the system, but you
                   dont seem to have the}%
       \typeout{'upmath' package installed. iaus.cls can take advantage
                 of these fonts,^^Jif you use 'upmath' package.^^J}%
      }
  \else
  \fi

  \checkfont{msam10}
  \iffontfound
    \IfFileExists{amssymb.sty}
      {\typeout{^^JFound AMS Symbol fonts on the system, using the
                'amssymb' package.^^J}%
       \usepackage{amssymb}%

      }{}
  \fi

  \IfFileExists{amsbsy.sty}
    {\typeout{^^JFound the 'amsbsy' package on the system, using it.^^J}%
     \usepackage{amsbsy}}
    {}

\usepackage{psfig}
\def\gsim{\lower 2pt \hbox{$\, \buildrel {\scriptstyle >}\over
{\scriptstyle \sim}\,$}}
\def\lsim{\lower 2pt \hbox{$\, \buildrel {\scriptstyle <}\over
{\scriptstyle \sim}\,$}}
\def\rosat{{\sl ROSAT}}

\def\chandra{{\sl Chandra}}

\newsavebox{\astrutbox}
\sbox{\astrutbox}{\rule[-5pt]{0pt}{20pt}}

\title[Outskirts of Galaxy Clusters: intense life in the suburbs]
      {Galaxy Cluster Formation from the Large-scale Structure: A Case Study of the Abell 2125 Complex at z=0.247}

\author[Q. D. Wang {\it et al.\/}]%
{Q. Daniel Wang$^1$, Frazer Owen$^2$,
Michael Ledlow$^3$, and William Keel$^4$}

\affiliation{$^1$Department of Astronomy, University of Massachusetts, 
Amherst, MA, USA\\[\affilskip]
$^2$National Radio Astronomy Observatory, Scorro, NM, USA\\[\affilskip]
$^3$Gemini Observatory, La Serena, Chile\\[\affilskip]
$^4$Department of Astronomy, University of Alabama, Tuscaloosa, AL, USA}

\pubyear{2004}
\volume{195}
\pagerange{1--8}
\date{?? and in revised form ??}
\jname{Outskirts of Galaxy Clusters: intense life in the suburbs}
\editors{A. Diaferio, ed.}
\begin{document}

\maketitle

\begin{abstract}
The structure of the universe is believed to have formed by clustering
hierarchically from small to large scales. Much of this evolution
occurs very slowly but at a few special times more, rapid, violent activity
may occur as major subunits collide at high velocities.  The Abell 2125
complex (z=0.247) appears to be undergoing such an event as shown by modeling of the
optical velocity field and by the detection with the VLA of an unusually large
number of associated radio active galaxies. We present an
80 ksec {\sl Chandra} imaging of Abell 2125, together with
extensive complementary multi-wavelength data. We show direct evidence
for galaxy transformation and destruction during the cluster formation.
The {\sl Chandra} data unambiguously
separate the X-ray contributions from discrete sources and large-scale
diffuse gas in the Abell 2125 complex, which consists of various
merging clusters/groups of galaxies and low-surface brightness emission.
This enables us to study processes
affecting galaxy evolution during this special time from scales of Mpc down
to a few kpc. The overall level of activity
plus the special time for the cluster-cluster merger suggests that an
important phase of galaxy evolution can take place during such events.
\end{abstract}

\firstsection 
\section{Introduction}
        We have identified a large-scale hierarchical complex 
(Fig.\ 1  left panel)
that is well-suited for investigating the structure formation process and
the galaxy/environment interactions. Revealed in a survey of 
10 Butcher \& Oemler clusters observed with the \rosat\ PSPC, this complex 
contains various X-ray-emitting features, which are associated with
concentrations of optical and radio galaxies (Fig.\ 1; 
Wang, Connolly, \& Brunner 1997; Wang, Owen, \& Ledlow 2004). 
The overall optical galaxy concentration of the region
was classified previously as a cluster Abell 2125 (richness 4). 
In addition to its large blue galaxy 
fraction ($\sim 20\%$), Abell 2125 also contains a unusually high number of
radio sources, a factor of a few higher than those in
more-or-less relaxed clusters at about the same redshifts.
The \rosat\ image and follow-up optical observations 
have further shown that the complex contains three well-defined 
X-ray bright clusters (Wang et al. 1997). In addition, substantial amounts of 
unresolved low-surface brightness X-ray emission (LSBXE) are also present.
The overall size of the complex is at least $\sim 5$~Mpc (assuming
the now standard $\Lambda$CDM cosmology).  
The complex thus represents an X-ray-bright hierarchical filamentary 
superstructure, as predicted by numerical simulations of the structure 
formation (e.g.,  Cen \&  Ostriker 1996). 


\begin{figure}
\centerline{
}
\caption{\protect\footnotesize
\rosat\  PSPC X-ray image of the Abell 2125 complex in the 
0.5-2~keV band (left panel; Wang et al. 1997). Point-like X-ray sources detected in the
image have been excised.  Spectroscopically confirmed, radio detected members of 
the Abell 2125 complex (Owen et al. 2004a) are marked by {\sl pluses}. The box 
outlines the field covered by our \chandra\ ACIS-I observation shown in the right panel, where the intensity image is in the 0.5-2 keV band. {\sl Squares} mark the positions of complex member 
galaxies.}
\label{fig1}
\end{figure}

We have obtained an 80 ksec \chandra\ observation that covers part of 
the Abell 2125 complex (Fig. 1, right panel) 
to characterize its detailed X-ray properties and to study the 
interplay between galaxies with their environments.
The high spatial resolution of \chandra\
enables us to detect point-like sources down to a limit of 
$\sim 1 \times 10^{-15}$ ${\rm~erg~cm^{-2}~s^{-1}}$ in the 0.5-8 keV band and 
to examine diffuse X-ray structures down 
to a scale of $\sim 3.8$~kpc ($\sim 1^{\prime\prime}$). Here we 
summarize results from the analysis of the data and 
from the comparison with extensive multi-wavelength observations of
the complex.

\section{New Results}

We detect a total of 99 discrete sources in the field. Ten of these sources
are identified with optical members of the complex. A statistical
analysis shows that few (if any) of the remaining unidentified sources 
are likely to be members of the complex.
The diffuse emission is substantially
softer than  typical discrete sources. There are substructures in the diffuse
X-ray emission on all scales.
The large-scale diffuse X-ray emission is correlated 
well with identified complex member galaxies (e.g., Fig. 1 right panel).
Two relatively prominent concentrations are sampled by the ACIS-I 
observation: the Abell 2125 cluster and the southwest LSBXE patch. 
The redshifts obtained from our X-ray spectra of these features are consistent 
the optical value ($z = 0.247$) of the complex.

\begin{figure}
\centerline{
}
\caption{\protect\footnotesize
{\sl HST} WFPC-2 V-band images of a field in the LSBXE and overlaid
X-ray contours in the the 0.5-2 keV band (left panel) and the 2-8 keV band (right panel). }
\label{fig17}
\end{figure}
 
\begin{figure}
\unitlength1.0cm
\caption{\protect\footnotesize
ACIS-I intensity images of the central cluster 
in the Kitt Peak 4-m mosaic V-band (red)
as well as in the 0.5 - 2 keV (green) and 2 - 8 keV  (blue) bands (left panel);
(right panel) a close-up of C153 in an {\sl HST} WFPC-2 V-band (red) and its diffuse X-ray trail.
}
\label{fig3}
\end{figure}

\subsection{Nature of the LSBXE}

The LSBXE is distinctly different from 
the central cluster. Even the relatively X-ray-bright southwest patch
does not seem to be centrally peaked, as would be expected 
from a more-or-less virialized intracluster medium (ICM). The kinematics
(Miller et al. 2004) indicates that the galaxies 
in this patch are probably loosely bound, although 
their exact line-of-sight distribution is not clear.
Fig. 2  shows a sub-field of the LSBXE. There are three bright ellipticals
in the field; but most of other galaxies appear to be spirals. X-ray
emission from such galaxies are typically dominated by X-ray binaries and/or
AGNs and is therefore expected to have a relatively hard spectrum. 
But the diffuse X-ray enhancement 
associated with the galaxy concentration is soft, indicating 
an origin in diffuse hot gas around individual galaxies,
in groups of galaxies, and/or in the intergroup IGM. 

The heavy element abundance in the LSBXE 
($< 10\%$ solar) appears to be substantially lower 
than that in the ICM of the central cluster
($\sim 1/4$ solar), although the abundance may be 
underestimated in the LSBXE if it contains multiple temperature components.
If the low metallicity obtained from the single temperature plasma fit is 
real, there may be two possible explanations: 1) Metals (mainly irons) are 
largely locked in dust grains within or 
around galaxy groups; 
2) The metals are still bound in individual galaxies. 
Dust grains may survive, and may even not mix well with, 
the X-ray-emitting gas in a 
quiescent, low density environment. But the mixing 
should likely occur during or after the merger of a galaxy or 
group with a cluster. The dust grains would then be 
destroyed rapidly by sputtering in the ICM of high density and temperature.
This may explain the metal abundance difference in the
X-ray-emitting gas between the LSBXE and
the central cluster.

\subsection{Dynamical State of the Central Cluster}

The central cluster (Abell 2125; Fig. 1) 
represents the strongest enhancement of the diffuse X-ray
emission in the complex. The relative values of our measured temperature 
(3 keV) and luminosity of the  ICM ($\sim 7.9 \times 10^{43} 
{\rm~ergs~s^{-1}}$ in the 0.5-2 keV band) are consistent with those 
for typical clusters. But, the presence of 
strong substructures 
in both the X-ray morphology and galaxy kinematics indicates that
the cluster is undergoing a merger.

The cluster also represents the densest concentration of galaxies in 
the Abell 2125 complex. Although most of the associated galaxies are 
quiet in radio, the brightest radio galaxies in the field
are found in the core of the cluster. While 
the galaxy C153 is a disturbed disk-like galaxy, the 
other three are all cD-like
ellipticals (Wang et al. 2004; Owen et al. 2004b). Interestingly, the two galaxies
(C153 and SW cD; Fig. 3)  associated with enhanced X-ray emission are both moving  
at high velocities ($\sim 1.5-1.8 \times 10^3 {\rm~km~s^{-1}}$) 
relative to the mean of the cluster. 

The exact relationship between the LSBXE and the central cluster is not
clear. The similar velocity centroids of the associated galaxy concentrations 
may indicate that they are about to merge along an 
axis that is perpendicular to the line of sight, although
there is little direct evidence from both X-ray and galaxy kinematics.

\subsection {Galaxy-Environment Interaction}

Most of 10 X-ray-bright member galaxies are not resolved by {\sl Chandra}
and are probably dominated by AGN activities.
Two of the X-ray sources are apparently resolved and are
associated with the disturbed disk-like galaxy C153 and 
the giant elliptical galaxy northwest of
the central Abell 2125 cluster (Fig. 3 left panel). 
As our source detection algorithms
are optimized to detect point-like sources, extended sources
such as the one associated with the SW cD 
(Fig. 3 left panel) are not detected. These extended soft X-ray enhancements
probably represent hot gas associated with individual massive galaxies or
groups of galaxies, which may have entered the cluster for the first time.
Outside the central cluster, the ambient density and relative velocity
are typically low and the ram-pressure stripping is
probably not important, at least for massive galaxies. The intergalactic
gas around galaxies may even cool fast enough to replenish 
the gas consumed for star formation (Bekki et al. 2002). 
As they are plugging into a cluster, the surrounding gas may then be 
compressed by the high ram-pressure of the ICM, resulting in
enhanced soft X-ray emission. The eventual stripping of the gas and dust 
may be important in both enriching the ICM and transforming the galaxies.

The most conspicuous soft X-ray feature in the
core of the central cluster is a ``trail'' 
apparently attached to the radio galaxy  C153 (Fig. 3 right panel). 
A similar, though shorter,
trail in the same direction is also seen in [OII] line emission 
(Owen et al. 2004b). 
C153 probably represents an extreme case of the ram-pressure stripping.
The observed steep X-ray spectral 
characteristics, as evidenced by the complete absence of the trail 
in the 2-8 keV band, suggests
a thermal gas temperature of $\sim 0.7$ keV, although 
the counting statistics of the data is not sufficient to give a tight 
constraint. We estimate the total luminosity of the trail as $\sim 5
\times 10^{41} {\rm~ergs~s^{-1}}$ in the 0.5-2 keV band.

\subsection{Abell 2125 as a Large-scale Hierarchical Complex}

This distinct complex of galaxies and hot gas shows several
remarkable characteristics:

\begin{figure}
\unitlength1.0cm
\centerline{
\psfig{figure=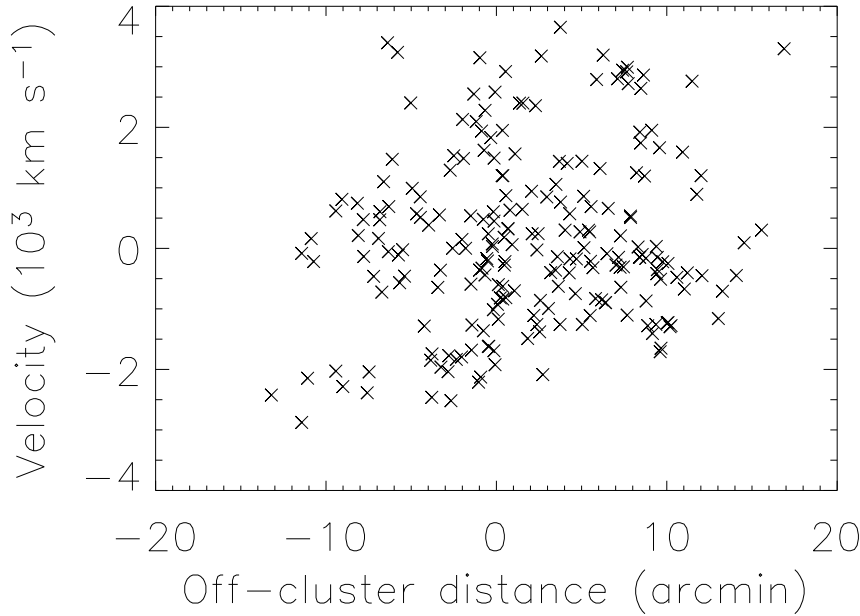,height=3.8truein,angle=0, clip=}
}
\caption{\protect\footnotesize
Radial velocities of galaxies versus the distance from the core cluster
along the major axis ($\sim 40^\circ$ east from the north) of the 
Abell 2125 complex. 
}
\label{fig7}
\end{figure}

\begin{itemize}
\item an unusually large galaxy velocity dispersion of $\sim 1.1 \times 10^3 
{\rm~km~s^{-1}}$, with multiple velocity components over a scale of $\sim 5$ Mpc (e.g., Fig. 4);

\item an exceptionally large fraction of radio galaxies, which are located
primarily outside the central Abell 2125 cluster (Owen et al. 2004a);

\item the presence of multiple X-ray-emitting clusters and LSBXE 
features, each with substantially lower luminosity and temperature
than expected from the overall galaxy richness and 
large velocity dispersion of the complex (Wang et al. 2004; Wang et al. 1997).
\end{itemize}

Comparisons with various numerical simulations 
indicate that the complex represents a projection of multiple components, 
which might be in a process
of merging with each other (Miller et al. 2004; Owen et al. 2004a). 
Projection effects, together with enhanced
activities during this process, may explain the observed characteristics.

\end{document}